\documentstyle[12pt]{article}
\def\ll{\label}
\def\re{\ref}
\def\c{\cite}

\def\r1{(\ref{$1})}

\def\ba{\begin{array}{c}}

\def\ea{\end{array}}

\def\de{\delta}

\def\l{\left}
\def\l({\left(}
\def\r){\right)}
\def\r{\right}

\def\be{\begin{equation}}
\def\bc{\begin{center}}
\def\ec{\end{center}}
\def\ee{\end{equation}}
\def\ed{\end{document}}
\def\bea{\begin{eqnarray}}
\def\eea{\end{eqnarray}}

\def\d{\partial}
\begin{document} 
\begin{center}
{\Large\bf
 Exact solution of a many body problem with nearest and next-nearest
neighbour interactions } \vspace{1.5cm} \\
{\large {\sc
B. Basu-Mallick \footnote {email:biru@tnp.saha.ernet.in} and
Anjan Kundu \footnote {email: anjan@tnp.saha.ernet.in}}}
\vspace{0.3cm}\\{\it
 Theory Group,
Saha Institute of Nuclear Physics,\\ 1/AF Bidhan Nagar,
 Calcutta 700 064, India.}\\
 \end{center}

\vskip 1 cm

\begin{abstract} 
{Recently  a partially solvable many-body problem with nearest and
next-nearest neighbour interactions is proposed \c{khare}. We show that by
adding a suitably chosen momentum dependent nearest neighbour interaction,
such a model can be converted into an integrable system with Lax operator
formulation and related conserved quantities. We also solve the eigenvalue
problem for the model exactly and as a byproduct obtain some identities
involving associated Laguerre polynomials.}
\end{abstract}

\smallskip


It is well known that  random matrix models have deep connection with 
some exactly solvable many-body  systems
 having  long-range interactions, namely 
Calogero \c{calogero} and Sutherland  model \c{suther}. 
However recently a new type of random matrix model, known as short-range 
Dyson model, has been introduced, where only the
nearest neighbour energy levels interact with each other
\c{jain98,bogo99}.
For finding out the corresponding many-body system, Calogero and Sutherland
like models with nearest  and next-nearest neighbour interactions have been
proposed recently \c{khare}. 
Though many interesting results, e.g.  construction of static correlation
functions, partial solution of the energy eigenvalue problem, existence
of off-diagonal long-range order etc. have been obtained, the integrability 
and  complete  solvability
of these models could not be established. 
 The Hamiltonian of such  Calogero like
model as considered in \c{khare,khare1}  is given by 
\be
 H ~=~ -  \, {1\over 2} \sum_{j=1}^N
 {\d^2 \over \d x_j^2}  + ~ {\omega^2\over 2} \,  \sum_{j=1}^N x_j^2
   ~+~  \nu (\nu -1 ) \,  \sum_{j=1}^{N-1} {1 \over (x_j -x_{j+1})^2}  
~ - \nu^2  \,  \sum_{j=2}^{N-1} {1 \over (x_{j-1} -x_{j}) (x_j -x_{j+1}) }.  
  \ll{hc} 
\ee

Our aim here is to propose an integrable and completely solvable 
variant of the above  model
by suitably redefining its coupling constants and  
 introducing an additional momentum dependent nearest neighbour
interaction.  
The Hamiltonian of this model may be given as 
\bea
 H ~=~ -  \, {1\over 2} \sum_{j=1}^N
 {\d^2 \over \d x_j^2} && + ~ {\omega^2\over 2} \,  \sum_{j=1}^N x_j^2
   ~+~ g_1  \,  \sum_{j=1}^{N-1} {1 \over (x_j -x_{j+1})^2}  
  \nonumber \\ && +
~ g_2  \,  \sum_{j=2}^{N-1} {1 \over (x_{j-1} -x_{j}) (x_j -x_{j+1}) } ~+~ 
   g_3  \, \sum_{j=1}^N  f_j { \d \over \d x_j }   ,
  \ll{h} 
\eea
 where the last term represents 
a momentum dependent interaction with
\be
f_j ~=~ (1-\delta_{1j}) { 1\over x_j - x_{j-1} }
 + (1-\delta_{N,j}) { 1\over x_j - x_{j+1} }
  \ll{fk} 
\ee
and the coupling constants are parametrised as 
\be 
g_1 = \nu (i + \nu) ,~~ g_2 = - \nu^2 , ~~g_3 = -i  \nu .
\ll{case1}
\ee
It is easy to see that  for  real parameter  $\nu$ 
the  Hamiltonian (\ref{h}) becomes  hermitian.
It  has been   shown  recently  \c{bk}
 that the original Calogero model with long-range 
 momentum dependent interaction
  retains its integrability and complete solvability. 
Similarly for the present model (\ref{h}),
 involving only nearest 
neighbour momentum dependent interaction,   one can establish 
the integrability and the existence of 
 infinite number of conserved quantities.
For this purpose one needs 
the associated   Lax operator,  
which may be expressed as
 a $N \times N$ matrix with  elements 
\be
L_{jk}= \ ( - i{\d \over \d x_j} + \nu f_j ) \  \de_{jk}. 
\ll{L}\ee
One may check that this  operator and Hamiltonian (\ref {h}) 
satisfy the Lax equation 
\be
[H,L^\pm]=\pm \omega L^\pm, ~~\mbox{where}~~ L^\pm_{jk}=L_{jk}\pm 
i \omega  x_j \de_{jk}. 
\ll{Le} \ee 
 Note that though the complementary Lax operator 
$M$ is absent in this   Lax  equation, it still  leads to 
a set of conserved quantities given by 
$I_m= \sum_{j,k}\left( (L^+ L^-)^m \right)_{jk}, \ m=1,2,3, \ldots ,$ 
ensuring the integrability of the system.
Evidently  Hamiltonian (\re{h})  is the first element of this set.

For solving the energy eigenvalue problem of  model  (\ref{hc}) and its
other variants,
  the authors  of
\c{khare,khare1}
 have adopted two different approaches. The first one, as described in
 \c{12}-\c{15},
 constructs the eigenfunctions 
by mapping  the model
to a system of free oscillators  through a suitable  similarity
transformation. In the second approach, pioneered by Calogero \c{calogero},
 the solution for eigenfunctions is 
sought through an ansatz in the factorised form:  $ P_k(x) \Phi(r^2),$    
where $P_k(x)$ is 
 a homogeneous polynomial in all coordinates $x=x_1,x_2, \ldots, x_N$
 and $ \Phi(r^2)$ is a function of $r^2=\sum_{i=1}^N x_i^2$.
Through both these methods    some particular but  exact
eigenvalue solutions were obtained in \c{khare,khare1}.

Due to the integrability property of our model (\ref{h}), we hope to 
get  the  complete  solution of  its   eigenvalue problem by adopting 
similar methods. Following the first approach, we find that 
there exists an unitary transformation
which reduces the Hamiltonian (\ref{h})
to a system of free oscillators:
\be 
\left ( \prod_{j=1}^{N-1} (x_j - x_{j+1}) \right )^{i \nu} ~ H ~
\left ( \prod_{j=1}^{N-1} (x_j - x_{j+1}) \right )^{-i \nu} ~=~H_{free} \, ,
\ll{hfree}
\ee
with 
\be
 H_{free} ~=~ -  {1\over 2} \sum_{j=1}^N
 {\d^2 \over \d x_j^2} + {\omega^2\over 2}  \sum_{j=1}^N x_j^2 \, .
  \ll{Hfree} 
\ee
As is well known, the eigenfunctions for the free oscillator model may be
given by
\be
\phi_{n_1, n_2, \cdots , n_N}~=~e^{- {\omega r^2 \over 2}} \,
 \prod_{j=1}^N H_{n_j}( {\sqrt \omega} x_j ) \, ,
\ll{phi}
\ee 
where $ r^2 ~=~ \sum_{j=1}^N x_j^2 $ and $ H_n(x)$  
is  the Hermite polynomial of degree $n.$
Therefore one can  immediately write the  eigenfunctions for our model 
(\ref{h}) as
\be
\psi_{n_1, n_2, \cdots , n_N}~=~
\left ( \prod_{j=1}^{N-1} (x_j - x_{j+1}) \right )^{-i \nu} ~
e^{- {\omega r^2 \over 2}} \, \prod_{j=1}^N H_{n_j}( {\sqrt \omega} x_j ) \, .
\ll{psi}
\ee 
  The corresponding eigenvalues 
are evidently same as that of the free oscillators:
\be 
E_{n_1, n_2, \cdots , n_N}~=~\omega \left( {N\over 2} + 
\sum_{j=1}^N n_j \right) \, .
\ll{efree}
\ee
The ground state wave function is clearly
 given by 
$\psi_{gr}~=~
\left ( \prod_{j=1}^{N-1} (x_j - x_{j+1}) \right )^{-i \nu} ~
e^{- {\omega r^2 \over 2}} $,  
 with energy eigenvalue $E_{gr} = {\omega N \over 2}$.
Thus we see that model (\ref{h}), which is obtained from (\ref{hc}) by 
adding a momentum dependent term and redefining  the coupling
constants, becomes exactly solvable  giving  very simple spectrum.
In a similar way, by adding suitably chosen momentum dependent interactions, 
the other variants of model (\ref{hc}) considered in \c{khare,khare1} can
also  be 
transformed  to integrable systems
 and solved exactly.

We should note here that the unitary transformation (\ref{hfree}), which
brings our Hamiltonian to the free oscillator model, does not
contain any singular terms. Consequently the eigenfunctions (\ref{psi})
form a complete set and the   spectrum coincides with that of 
{\it distinguishable} oscillators.
In contrast, the similarity transformations which map the
  standard Calogero model as well as (\ref{hc})
  into  the
free oscillator model contain singular terms.
In the case of Calogero model such singularities can be avoided 
 by symmetrising the  eigenfunctions of free oscillators
 in all coordinates \c{12}. As a result the
 eigenfunctions of Calogero model represent a complete set with the 
excitation spectrum becoming equivalent to that of  bosonic oscillators.
However in the case of  (\ref{hc}) such singularities can not be removed in
general  even by taking symmetrised eigenfunctions 
 and that makes  the completeness  of eigenfunctions difficult
to establish.

It may be  noted  that  
both (\ref{hc}) and our model (\ref{h})
reduce to   
 the  free oscillator model at  $\nu \to 0.$ 
Therefore it may be expected that a link should exist between the  solutions 
of these two models at $\nu \to 0 $ limit.
To examine this possibility we  follow the second  approach, as mentioned
above, 
for solving the model  (\ref{h}) and find that the
corresponding  eigenfunctions  
 can   be obtained  in the factorised form
\be 
\Phi_{n,k} ~=~ e^{- {\omega r^2 \over 2}} 
\left ( \prod_{j=1}^{N-1} (x_j - x_{j+1}) \right )^{-i \nu}
L_n^{{N\over 2} + k -1}(\omega r^2) \, P_k(x) \, .
\ll{phif}
\ee
Here $ P_k(x)$ is a homogeneous 
  polynomial of degree $k$ satisfying the
equation 
\be 
\sum_{j=1}^N \, {\d^2 \over \d x_j^2} \, P_k(x_1, x_2, \cdots , x_N)~=~0 \, ,
\ll{harmon}
\ee
and $L_n^{{N\over 2} + k -1}(\omega r^2) $
 is the associated Laguerre polynomial. 
The corresponding  eigenvalue would be 
$ E_{n,k} ~=~ \omega \left ( {N\over 2} + 2n +k  \right )$.
A class of trivial solutions of (\ref{harmon})  with $k \leq N$ 
is clearly given by 
\be
P_k(x) ~=~ x_{\alpha_1} x_{\alpha_2} \cdots x_{\alpha_k} \, ,
\ll{pkl}
\ee
where $\alpha_l$s are all different integers ranging from $1$ to $N$. 
For finding nontrivial solutions of (\ref{harmon}) we express 
$P_k(x)$ in a general form
\be
P_k(x) ~=~  \sum_{ r_1 + r_2 + \cdots + r_N = k} ~{ k! \over 
r_1!   r_2!  \cdots  r_k! }~f^k_{r_1, r_2, \cdots , r_N} ~
x_1^{r_1} x_2^{r_2} \cdots  x_N^{r_N},
\ll{pkx}
\ee
where $f^k_{r_1, r_2, \cdots , r_N}$
 are yet unknown coefficients and the summation variables $r_j$ 
are nonnegative integers. The equation (\ref{harmon})
however puts constraints on these coefficients as
\be 
\sum_{j=1}^N ~ f^k_{r_1, \, r_2, \, \cdots , \, r_{j-1} , ~ r_j 
\, + \, 2 , \, r_{j+1}, \, r_{j+2}, \cdots , \, r_N } ~=~0 \, ,
\ll{f0}
\ee
where
  $\{r_j\}$ represents any partition of $k-2$ satisfying $\sum_{j=1}^N
 r_j=k-2$. 
Such constraints  are evidently different for
different choices of $\{r_j\}$.
However since the number of constraints is less than the number of unknown
coefficients, they  can not be fixed  uniquely unless one imposes
some extra  relationship among them. Therefore  the corresponding
polynomial (\ref{pkx}) 
may not be symmetric in all its coordinates in general.
 For example for  $k=2$ one gets the solution 
\be 
P_2(x) ~=~ a_1 x_1^2 + a_2 x_2^2 + \cdots + a_N x_N^2 , ~~~
\mbox {with} ~\sum_{j=1}^N a_j = 0,
\ll{p2}
\ee
which can not be presented in a symmetric form.
On the other hand for higher $k$ one may obtain symmetric polynomials
as solutions of (\ref{f0}).
For $k=3$  such a solution is explicitly  given by
\be 
P_3(x)~=~a \, \sum_{j=1}^N x_j^3  \, + \, b \, \sum_{j \neq k} x_j^2 x_k \, ,
\ll{p3}
\ee
where $ 3a + (N-1)b = 0.$  For $k=4$ we get however 
two independent solutions of the form
\be 
P_4(x)~=~a \, \sum_{j=1}^N x_j^4  \, +\, b \, \sum_{j \neq k} x_j^2 x_k^2 \, ,
\ll{p41}
\ee
with the constraint $ 3a + (N-1)b = 0,$ and 
\be 
P_4(x)~=~c \, \sum_{j \neq k} x_j^3 x_k
  \, + \, d \, \sum_{j \neq k \neq l \neq j} x_j^2 x_k x_l  \, ,
\ll{p42}
\ee
with $ 3c + (N-2)d = 0$.

We may compare now the above solutions at $\nu \to
0$ with those of (\ref{hc}), as obtained in \c{khare}, 
 at the same limit.
 For the case $k=2$ while we get two solutions with  $P_2(x)$ taking 
 the trivial form (\ref{pkl}) as well as the nontrivial one (\ref{p2}),
 \c{khare} gives only one solution with  trivial form of $P_2(x)$.
 For $k=3$ we get again two independent solutions corresponding to 
(\ref{p3}) and
(\ref{pkl}), while in \c{khare} one finds only 
 a particular combination
of them. Similarly for $k=4$ we get three independent solutions
in contrast to  a single  solution of \c{khare}. 
This picture prevails also for higher $k$.

Finally we would like to comment on some useful identities, which may be
 derived by comparing the eigenfunctions (\ref{psi})
and (\ref{phif}) obtained  through two different approaches.
Expressing  each of the solutions   
 (\ref{phif}) as some linear combinations of the complete set
  (\ref{psi}),  one would generate such an
identity  involving associated Laguerre and Hermite polynomials.
  By using this procedure along with the familiar conversion formulas like
$H_{2n}(x) = (-1)^n 2^{2n}n! L_n^{
- {1\over 2}}(x^2)$, $H_{2n+1}(x) = (-1)^n 2^{2n +1}n! x L_n^{
 {1\over 2}}(x^2)$, we find interestingly  that (\ref{phif}) with 
the trivial solution (\ref{pkl})  for $P_k(x)$ 
 leads to a well known summation formula \c{RG}
for associated Laguerre polynomials:
\be 
L_n^{{N\over 2} + k -1}(\omega r^2) ~=~ \sum_{ n_1 + n_2 + \cdots +
n_N = n} ~\prod_{r=1}^k L^{1\over 2}_{n_r} ( \omega x_r^2 ) 
~\prod_{r=k+1}^N L^{-{1\over 2}}_{n_r} ( \omega x_r^2 ) \, .
\ll{RG}\ee
 On the other hand, (\ref{phif}) with 
nontrivial solutions  for $P_k(x)$ leads to apparently new
identities involving associated Laguerre polynomials.
For example, the  solution (\ref{p2})  for $N=2$ giving  $P_2(x)
= x_1^2 - x_2^2$ results an identity of the form
\be 
L_n^2(\omega x_1^2 + \omega x_2^2 ) ~=~ {1 \over\omega (x_1^2 - x_2^2)} \,
 \sum_{s=0}^{n+1} \, (2s-n-1) 
 L^{-{1\over 2}}_{n-s+1} (\omega x_1^2) L^{-{1\over 2}}_{s}
 (\omega x_2^2) \, . 
\ll{ni2}\ee
We hope to find similar identities for higher values of $k$ and $N$ 
and report elsewhere \c{up}.

\end{document}